\documentclass[iop,numberedappendix,apj]{emulateapj}
\pdfoutput=1
\usepackage{amsmath,amssymb,epsf,graphicx,morefloats,natbib,ulem,url,xcolor}
\usepackage[colorlinks=true,linkcolor=blue,anchorcolor=blue,citecolor=blue,urlcolor=blue,draft=false]{hyperref}

\providecommand{\sorthelp}[1]{}

\usepackage{color}
\definecolor{DarkGreen}{rgb}{0.0, 0.5, 0.0}
\definecolor{purple}{rgb}{0.5, 0.0, 0.5}
\definecolor{red}{rgb}{1, 0.0, 0.0}
\definecolor{DarkRed}{rgb}{0.7, 0.0, 0.0}
\definecolor{green}{rgb}{0, 1.0, 0.0}

\newcommand{\xmm}{XMM-{\it Newton}}
\newcommand{\plckg}{PLCK~G147.3-16.6}
\newcommand{\sigT}{\mbox{$\sigma_{\mbox{\tiny T}}$}}
\newcommand{\mec}{\mbox{$m_{\mbox{\tiny e}} c^2$}}
\newcommand{\te}{\mbox{$T_{\mbox{\tiny e}}$}}
\newcommand{\Tx}{\mbox{$T_{\mbox{\tiny X}}$}}

\newcommand{\kB}{\mbox{$k_{\mbox{\tiny B}}$}}

\newcommand{\Ysphf}{\mbox{$Y_{\mbox{\scriptsize 500,sph}}$}}

\newcommand{\Pe}{\mbox{$P_{\mbox{\scriptsize e}}$}}
\newcommand{\DA}{\mbox{$D_{\mbox{\tiny A}}$}}
\newcommand{\LCDM}{\mbox{$\Lambda$CDM}}
\newcommand{\dene}{\mbox{$n_{\mbox{\tiny e}}$}}

\newcommand{\Sx}{\mbox{$S_{\mbox{\tiny X}}$}}

\newcommand{\comment}[1]{}

\shorttitle{GISMO High-Resolution SZ Imaging of PLCK~G147.3-16.6}
\shortauthors{Mroczkowski, Kov{\'a}cs, Bulbul et al.}

\begin{document}

\title{Resolving the Merging {\it Planck} Cluster PLCK~G147.3-16.6 with GISMO}

\author{
T.~Mroczkowski\altaffilmark{1,2},
A.~Kov{\'a}cs\altaffilmark{3,4},
E.~Bulbul\altaffilmark{5},
J.~Staguhn\altaffilmark{6,7},
D.~J.~Benford\altaffilmark{6},
T.~E.~Clarke\altaffilmark{2},
R.~J.~van~Weeren\altaffilmark{5},
H.~T.~Intema\altaffilmark{8},
S.~Randall\altaffilmark{5}
}
\slugcomment{}
\date{\today}

\affil{\altaffilmark{1}National Research Council Fellow, 
National Academy of Sciences}
\email{E-mail: anthony.mroczkowski.ctr@nrl.navy.mil}

\affil{\altaffilmark{2}U.S.\ Naval Research Laboratory, 4555 Overlook Ave SW, 
Washington, D.C.\ 20375, USA}

\affil{\altaffilmark{3}California Institute of Technology 301-17, 
1200 E California Blvd, Pasadena, CA 91125, USA}

\affil{\altaffilmark{4}Institute for Astrophysics, University of Minnesota,
116 Church St SE, Minneapolis, MN 55455, USA}

\affil{\altaffilmark{5}Harvard-Smithsonian Center for Astrophysics, 
60 Garden Street, Cambridge, MA 02138, USA}

\affil{\altaffilmark{6}Observational Cosmology Lab., Code 665, NASA at 
Goddard Space Flight Center, Greenbelt, MD 20771, USA}

\affil{\altaffilmark{7}Department of Physics \& Astronomy, Johns Hopkins 
University, Baltimore, MD, 21218, USA}

\affil{\altaffilmark{8}National Radio Astronomy Observatory, P.O.\ Box O,
  1003 Lopezville Road, Socorro, NM 87801-0387, USA}

\begin{abstract} 
The {\it Planck} satellite has recently completed an all-sky galaxy cluster
survey exploiting the thermal Sunyaev-Zel'dovich (SZ) effect to locate some of the 
most massive systems observable.  
With a median redshift of $\left<z\right>=0.22$, the clusters found
by {\it Planck} at $z>0.3$ are proving to be exceptionally massive
and/or disturbed systems.
One notable {\it Planck} discovery at $z=0.645$, PLCK~G147.3-16.6, has an elongated 
core and hosts a radio halo, indicating it is likely in the process of merging.
We present a 16\arcsec\!\!.5 resolution SZ observation of this high-$z$ merger
using the Goddard-IRAM Superconducting 2-Millimeter Observer (GISMO), and compare 
it to X-ray follow-up observations with XMM-{\it Newton}. We find the
SZ pressure substructure is offset from the core components 
seen in X-ray. We interpret this as possible line of sight temperature or 
density substructure due to the on-going merger.
\end{abstract}


\section{Introduction}\label{sec:intro}

Forming from the largest fluctuations in the primordial matter power 
spectrum, galaxy clusters are among the most massive gravitationally-bound objects. 
Therefore, the distribution of clusters as a function of mass and redshift provides 
sensitive cosmological probes.  Surveys spanning the electromagnetic spectrum 
are planned or underway to catalog clusters across their formation history.
Recent efforts exploiting the redshift-independent
surface brightness of the Sunyaev-Zel'dovich effect (SZ; \citealt{sunyaev1972})
in particular have detected $\sim$1000 previously-unknown clusters \citep[see][]{fowler2010,
Carlstrom2011,Planck2011I,Planck2011ESZ,Hasselfield2013,Planck2013XXIX,Bleem2015}. 

The {\it Planck} satellite has completed the first 
all-sky cluster survey since ROSAT \citep[see e.g.][]{Romer1994,voges1999,bohringer2000}.
{\it Planck}, however, is not well-suited for the discovery
of high-$z$ systems, whose arcminute-scale SZ signals are heavily diluted inside {\it Planck}'s 
7.\!\arcmin3--9.\!\arcmin7 beams at the detecting 2 \& 3~mm bands. As such, {\it Planck} detects 
only the most prominent, rare systems at high-$z$.
The {\it Planck} \xmm\ cluster validation program \citep{Planck2013_XMMValidation} 
used the 15.5-month nominal 
survey data to identify likely cluster candidates and understand {\it Planck}'s selection
function. It suggests that the high-$z$ detections are likely dynamically-disturbed massive systems, which are far from 
being virialized and, on average, less X-ray luminous than X-ray selected clusters of the same mass.

Here we report high-significance 16\arcsec\!\!.5 resolution SZ observations 
of a disturbed cluster from the final {\it Planck} \xmm\ cluster validation program, 
imaging it with nearly $20\times$ better resolution than its original unresolved detection. 
These new data, from the Goddard-IRAM 2-Millimeter Observer (GISMO; \citealt{Staguhn2006}) 
on the 30-meter Institut de Radioastronomie Millim{\'e}trique (IRAM) Telescope\footnote{This 
work is based on observations carried out with the 30-meter IRAM Telescope. IRAM is 
supported by INSU/CNRS (France), MPG (Germany) and IGN (Spain).} on Pico Veleta, Spain, 
reveal complex pressure substructure in this merging system and underscore the power of subarcminute SZ follow-up. 

\begin{deluxetable*}{lccccc}
\tablewidth{0pt}
\tablecolumns{6}
\tablecaption{Observations}
\tablehead{
\underline{Observatory} & 
\underline{Date} & 
\underline{Project Code} & 
\multicolumn{2}{c}{\underline{Pointing (J2000)}} & 
\underline{Clean Exposure Time} \\
       &
       &
       &
  R.A. & 
  Dec. & 
(ksec)} 
\startdata 
GISMO 			& 7--9 Apr 2014 & 235-13     & 02:56:20.0 & +40:17:21.0 & 16.9 \\
GMRT            & 18 Jan 2013   & 23\_013    & 02:56:25.2 & +40:17:18.7 & 22 \\
\xmm\           & 27 Aug 2012   & 0693661601 & 02:56:23.8 & +40:17:28.0 & 41.7/42.1/32.9\tablenotemark{a} \\
\xmm\           & 22 Aug 2011   & 0679181301 & 02:56:25.3 & +40:17:18.7 & 15.4/15.7/8.6\tablenotemark{a}
\enddata
\label{tab:obs}
\tablecomments{Aim points and unflagged exposure times for the
 observations of \plckg\ included here.}
\tablenotetext{a}{Times for the MOS1, MOS2, and PN detectors of \xmm\ EPIC, respectively.}
\end{deluxetable*}

We summarize the known cluster properties in Section~\ref{sec:plckg147}, 
discuss the new observations in Section~\ref{sec:observations}, and present 
the results of our analysis in Section~\ref{sec:results}.
We adopt a \LCDM\ cosmology with $\Omega_M=0.3$, $\Omega_\Lambda=0.7$, and
$H_0=70\rm~km~s^{-1}~Mpc^{-1}$ throughout this paper.
At the redshift of \plckg\ \citep[$z=0.645$;][]{Planck2013_XMMValidation}, 
1\arcsec\ corresponds to 6.9~kpc.

\section{\plckg}\label{sec:plckg147}

\plckg\ is a massive cluster at $z$=0.645, discovered at a signal
to noise ratio $S/N$=4.41 in the nominal 15.5-month {\it Planck} mission.  
X-ray follow-up observations in the {\it Planck} \xmm\ validation program
\citep{Planck2013_XMMValidation} reveal an extended, 
double core morphology, while optical observations with Gemini show no cD 
galaxy dominating the cluster field.  
More recently, 610~MHz observations with the Giant 
Metrewave Radio Telescope (GMRT) found that \plckg\ hosts a 0.9~Mpc radio halo
\citep{vanweeren2014}, placing it among the highest redshift radio halos known. 
The (re-)acceleration processes that create radio halos are thought to
occur predominantly in mergers during and after the first core passage 
\citep{feretti2012,brunetti2014}, although at least one relaxed, cool-core 
cluster is known to host a radio halo \citep{Bonafede2014}. 
The disturbed X-ray morphology and lack of a dominant cD galaxy indicate that 
\plckg\ belongs to the traditional category of merging clusters hosting radio 
halos.
 
\section{Observations}\label{sec:observations}

In this section we describe the resolved GISMO SZ observations and new \xmm\ data 
presented in this work.
Table \ref{tab:obs} summarizes the observations presented here, including 
those from the GMRT reported in \citet{vanweeren2014} and the previous \xmm\
observations in \citet{Planck2013_XMMValidation}.

\subsection{GISMO}\label{sec:gismo}

We observed \plckg\ for a total of 4.7~hours (16.9~ksec) in April 2014 using
GISMO, an 8$\times$16 element array of transition edge sensors. 
From the 30-meter IRAM Telescope, 
GISMO provides a $1.\!\arcmin8$$\times$$3.\!\arcmin8$ instantaneous field of view 
with 16\arcsec\!\!.5 resolution at 150~GHz (2~mm).
The GISMO data were reduced with 
{\it CRUSH}\footnote{
\url{http://www.submm.caltech.edu/~sharc/crush}} 
\citep[ver.~2.22-1;][]{Kovacs2008}, which was optimized to recover
extended, diffuse signals from the atmosphere-dominated bolometer data.

We estimate a total calibration uncertainty $\sim$7--9\% due to the imperfect knowledge of the line of 
sight opacities, based on repeated observations of Mars and Uranus \citep[see][for a 
detailed description of the absolute calibration]{Staguhn2014}.

\begin{figure*}[bth]
\begin{center}
 \includegraphics[height=1.88in]{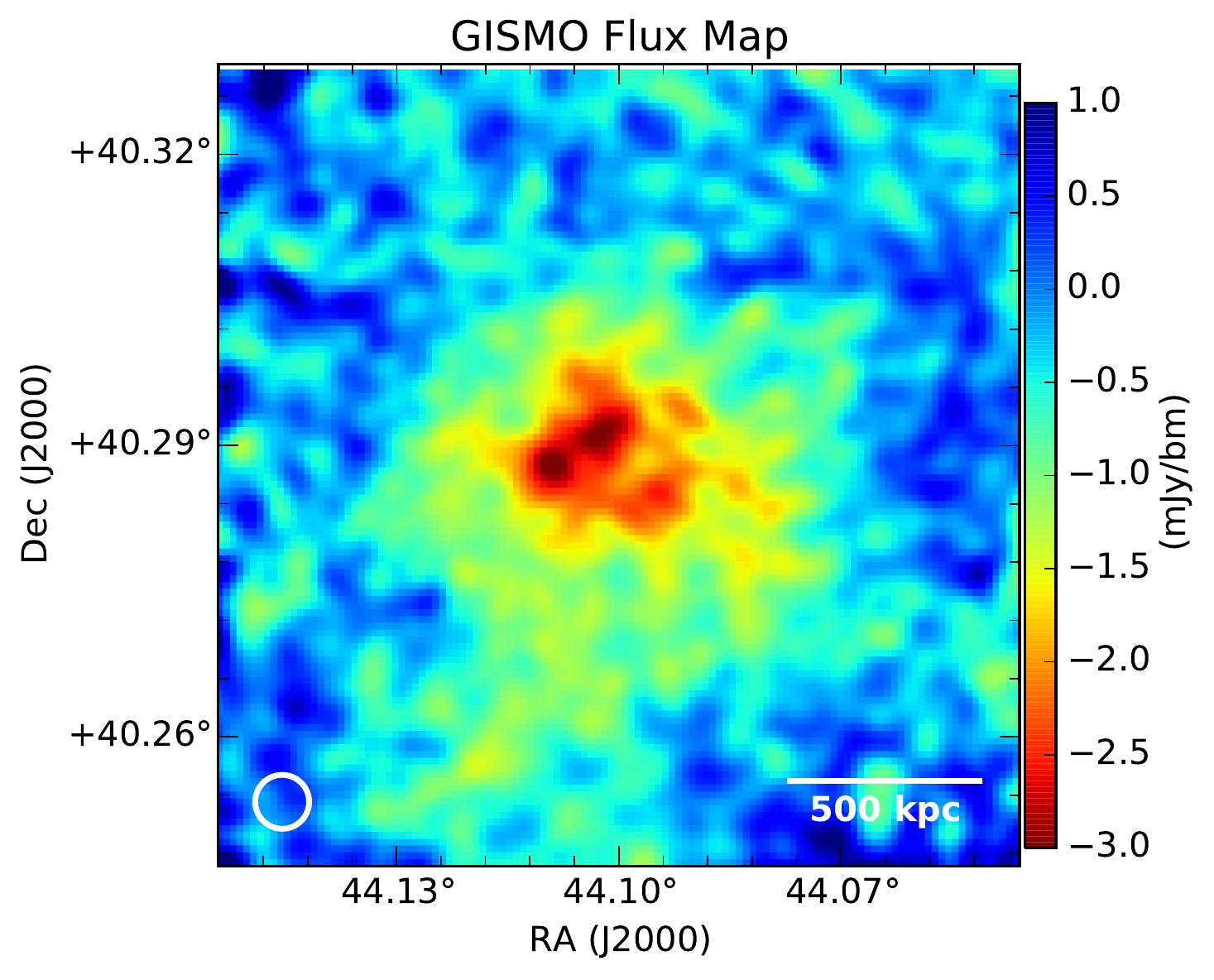}
 \includegraphics[height=1.88in]{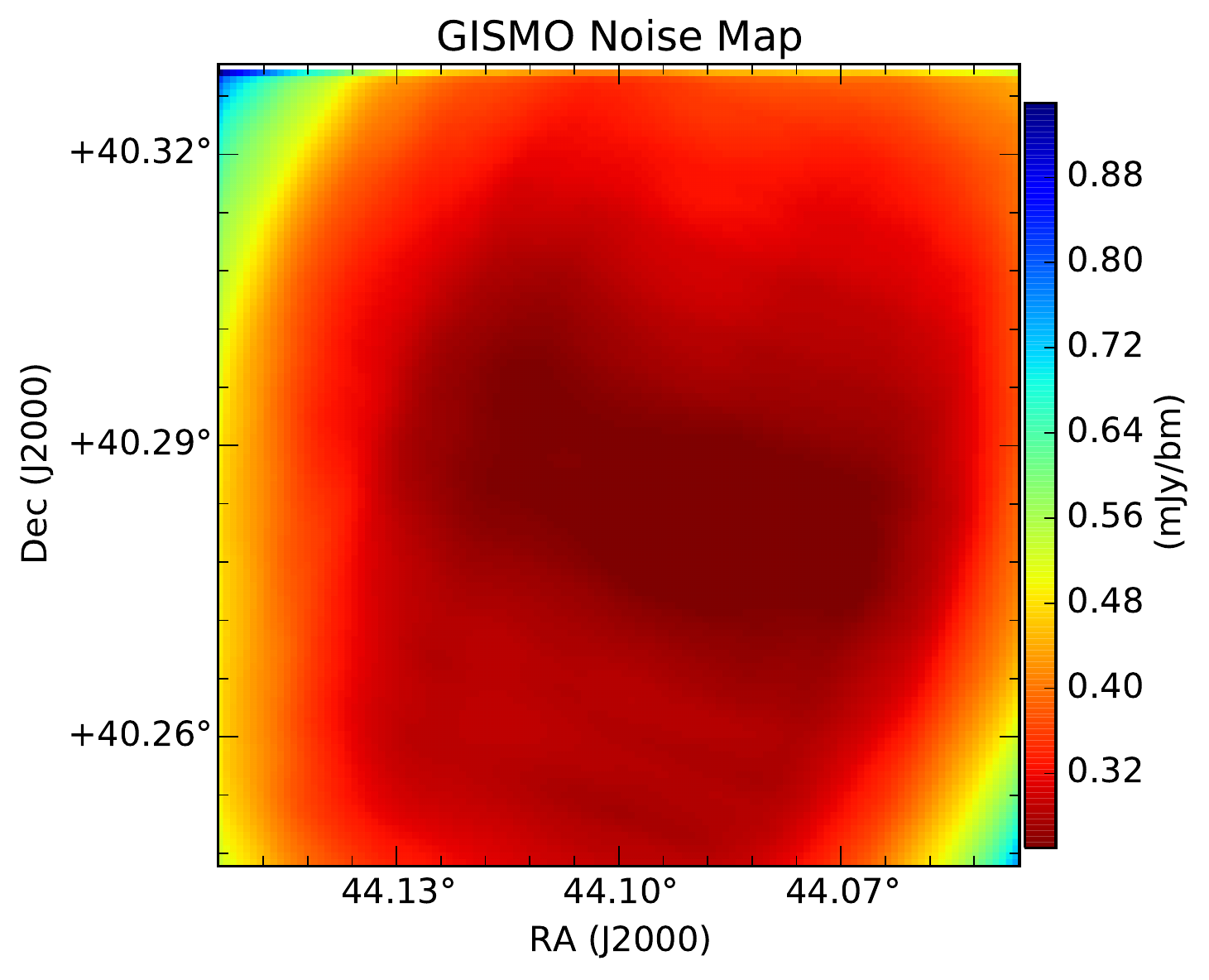}
 \includegraphics[height=1.88in]{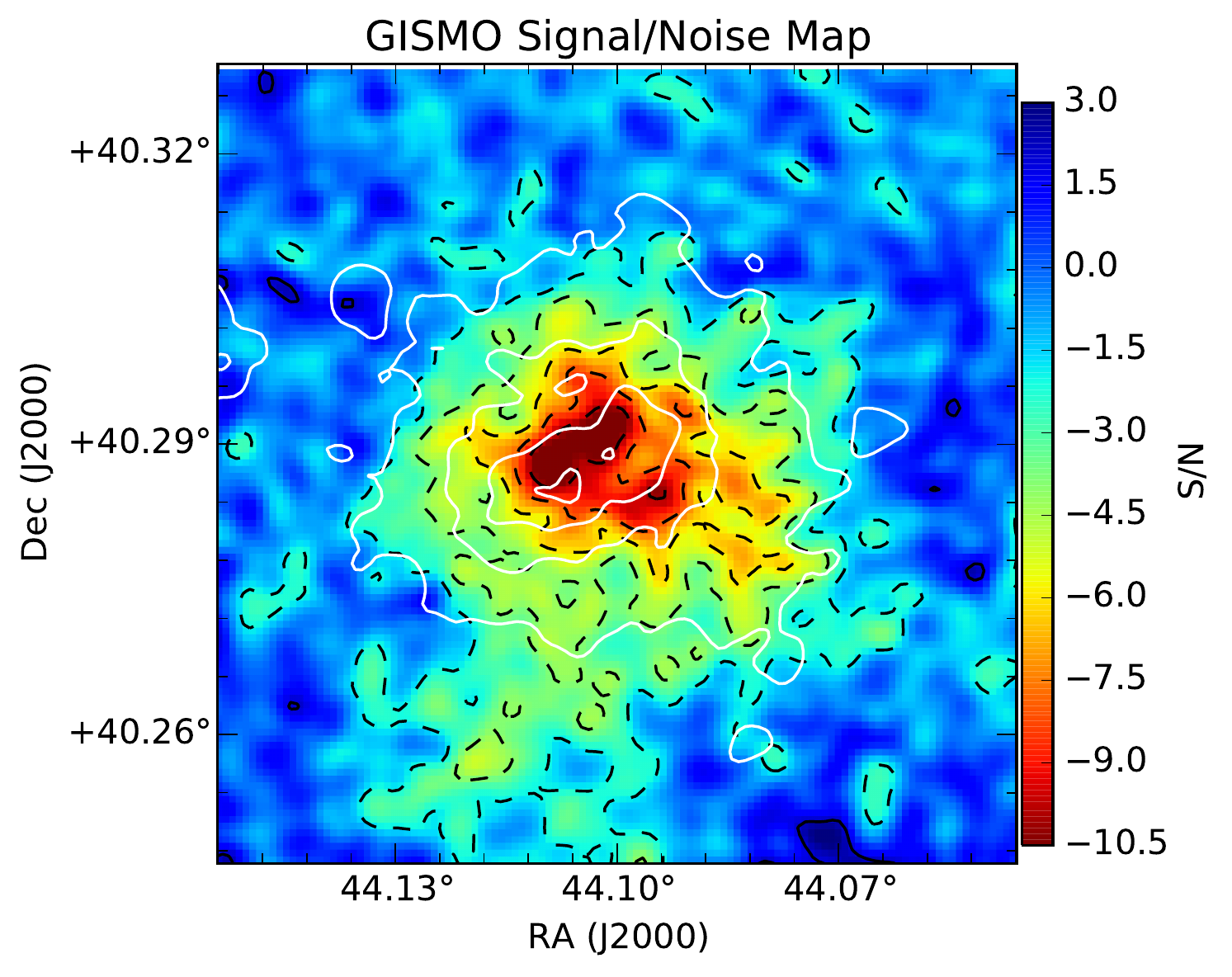}
\end{center}
\caption{ 
  {\bf Left:} Deconvolved GISMO flux map, smoothed by a 12\arcsec Gaussian to 
  a resolution of 19.\!\arcsec2 (depicted in the lower left corner), showing the 
  SZ decrement (mJy/bm) toward \plckg. 
  {\bf Middle:} GISMO noise map with nearly uniform coverage in the region where 
  the cluster is best detected. 
  {\bf Right:} Signal-to-noise ($S/N$) map (the ratio of left and middle images), 
  with contours at $S/N=[4,2,-2,-4,-6,-8,-10]$ overlaid in black. X-ray contours from 
  Figure~\ref{fig:cluster_xray} (left panel) are overlaid in white, starting at 3-$\sigma$ and 
  spaced at 3-$\sigma$ intervals.
  \label{fig:cluster}}
\end{figure*}

We deconvolved the resulting image with the measured point-source response of the reduction process;
i.e., we divide the Fourier transform of the map by the 2-D transfer function, and back-transform. 
The resulting deconvolved image is shown in Figure~\ref{fig:cluster} (left).

The cluster field was observed using a combination of 3--5\arcmin\ alt-azimuthal Lissajous patterns,
yielding a median noise $\approx$0.3~mJy/bm within the central 4\arcmin\ diameter, and coverage 
extending to approximately $6\arcmin\!\!.3$$\times$$8\arcmin\!\!.3$ area overall.

The noise in each map pixel was propagated from noise measured in the residual detector 
timestreams. Non-white (covariant) features, such as residual $1/f$, is spatially invariant, 
and hence fully captured by an appropriate noise re-scaling, which we determined by the ratio 
of measured-to-expected deviation, $\left< (S_{i,j} / \sigma_{i,j})^2 \right>^{1/2}$, outside 
of the approximate cluster center (at $r$$>$1.\!\arcmin5). 
The resulting noise map, shown in Figure~\ref{fig:cluster} (middle), provides a fair measure of 
the true map noise for GISMO, with no apparent transient noise.  We find the peak SZ decrement 
at $>$10-$\sigma$ significance, and the overall detection is significant at $>$3-$\sigma$ in
every beam within the central $\gtrsim 2\arcmin$ of the map.

\begin{figure}
\begin{center}
	\includegraphics[width=3.45in]{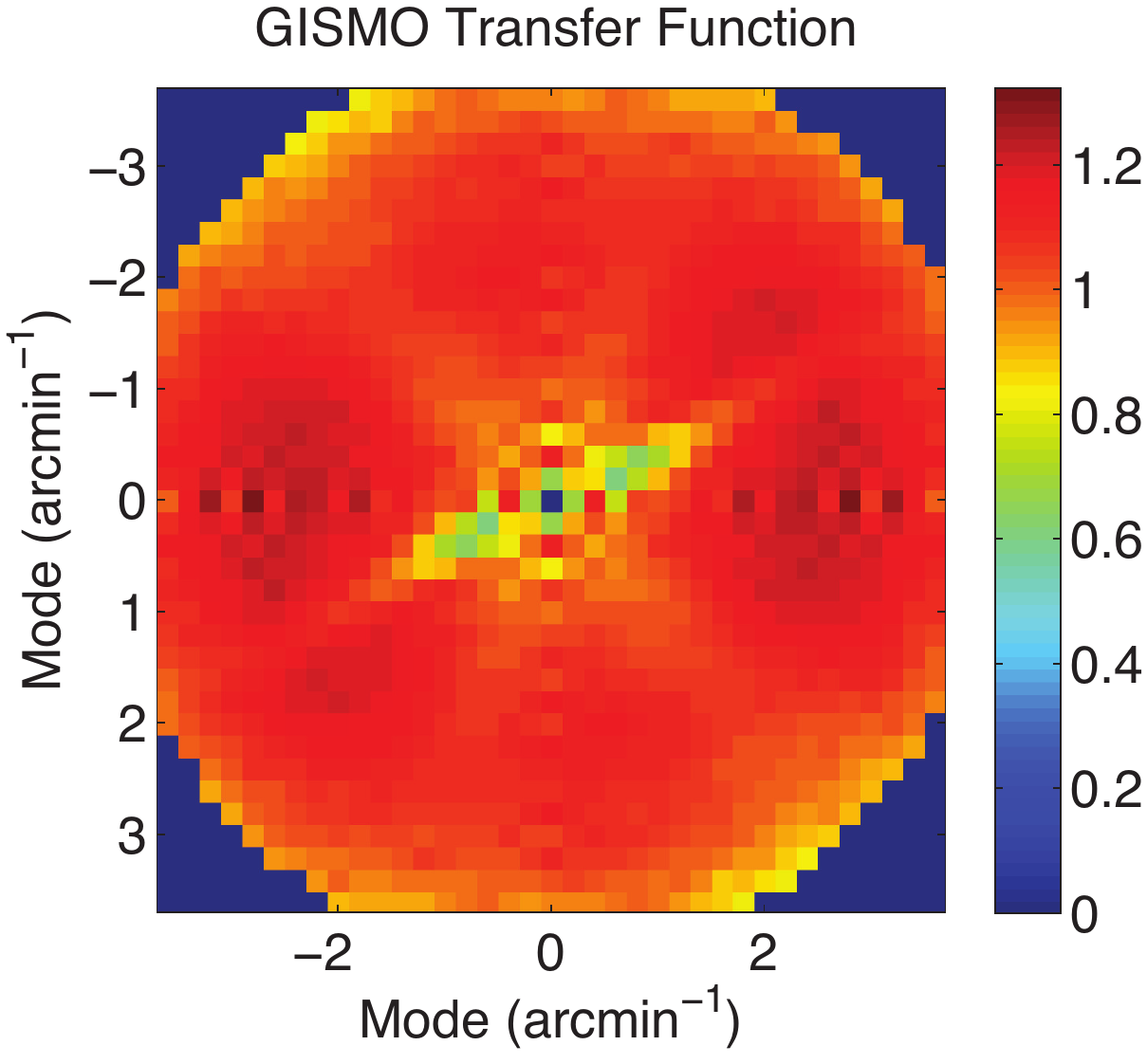}\\
	\includegraphics[width=3.3in]{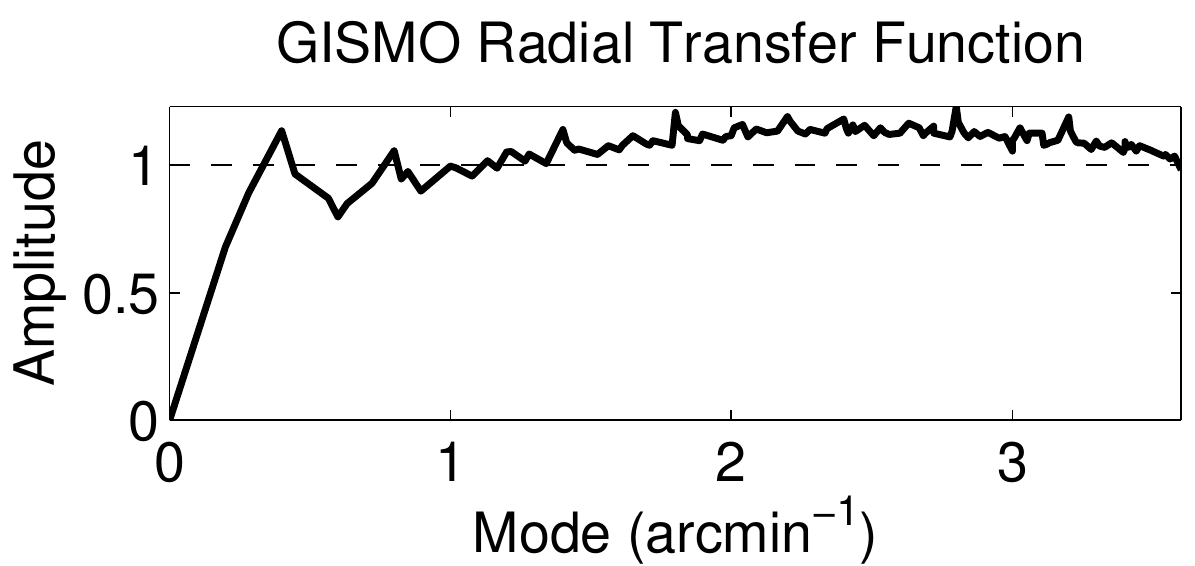}
\end{center}
\caption{ {\bf Upper:} 2D transfer function used for deconvolution 
	of the \plckg\ GISMO observation. Note: the corner values at 
	radii $>3.2$~arcmin$^{-1}$ are indeterminate.
	 {\bf Lower:} Radial profile of the above transfer function.
	  \label{fig:xfer}}
\end{figure}

The two-dimensional transfer function of our GISMO data, shown in Figure~\ref{fig:xfer}, 
was obtained by inserting a faint point-like test source into jackknife realizations, which 
are reduced the same way as the cluster. Thus, we ensure that the test dataset has the same 
noise properties as the actual dataset, and therefore the test source undergoes the 
same filtering steps as our cluster, even with adaptive pipeline steps such as noise whitening.
We averaged the response over 100 jackknife realizations to suppress the low-level
sky-noise present in the individual realizations. The transfer function, obtained as the 
ratio of the observed 2D spatial spectrum of the response to the underlying spectrum of the 
test source, characterizes the pipeline's response to arbirtrary structures.

Our transfer function is not circularly symmetric due to common mode subtraction along 
various correlated detector groups. Because {\it CRUSH} normalizes maps to preserve 
point-source peak fluxes (i.e.\ to keep the point source weighted mean response unity by definition), 
the response in the raw GISMO map diminishes gradually from above unity at the short 
spatial scales (20\arcsec; 3~arcmin$^{-1}$), to $<$1 at scales 
$\gtrsim$2\arcmin, which is comparable to the instantaneous field of view.
The azimuthally-averaged transfer function is also shown in Figure~\ref{fig:xfer} (lower panel).

The raw map deconvolved by the transfer function provides an accurate 
representation of the underlying 2-mm flux distribution of \plckg\ up to the 5\arcmin\ 
scales shown. The reduction and deconvolution algorithm was tested on both a simulated point source and a
simulated cluster model, and it accurately reproduced the expected fluxes and profiles for both 
(see Fig.~\ref{fig:sz_profiles}). The zero level of the deconvolved map is estimated using the mean flux level 
outside the cluster decrement (at radii $R>1.5\arcmin$; see Section~\ref{sec:results}).

\begin{figure*}[bth]
\begin{center}
 \includegraphics[height=1.88in]{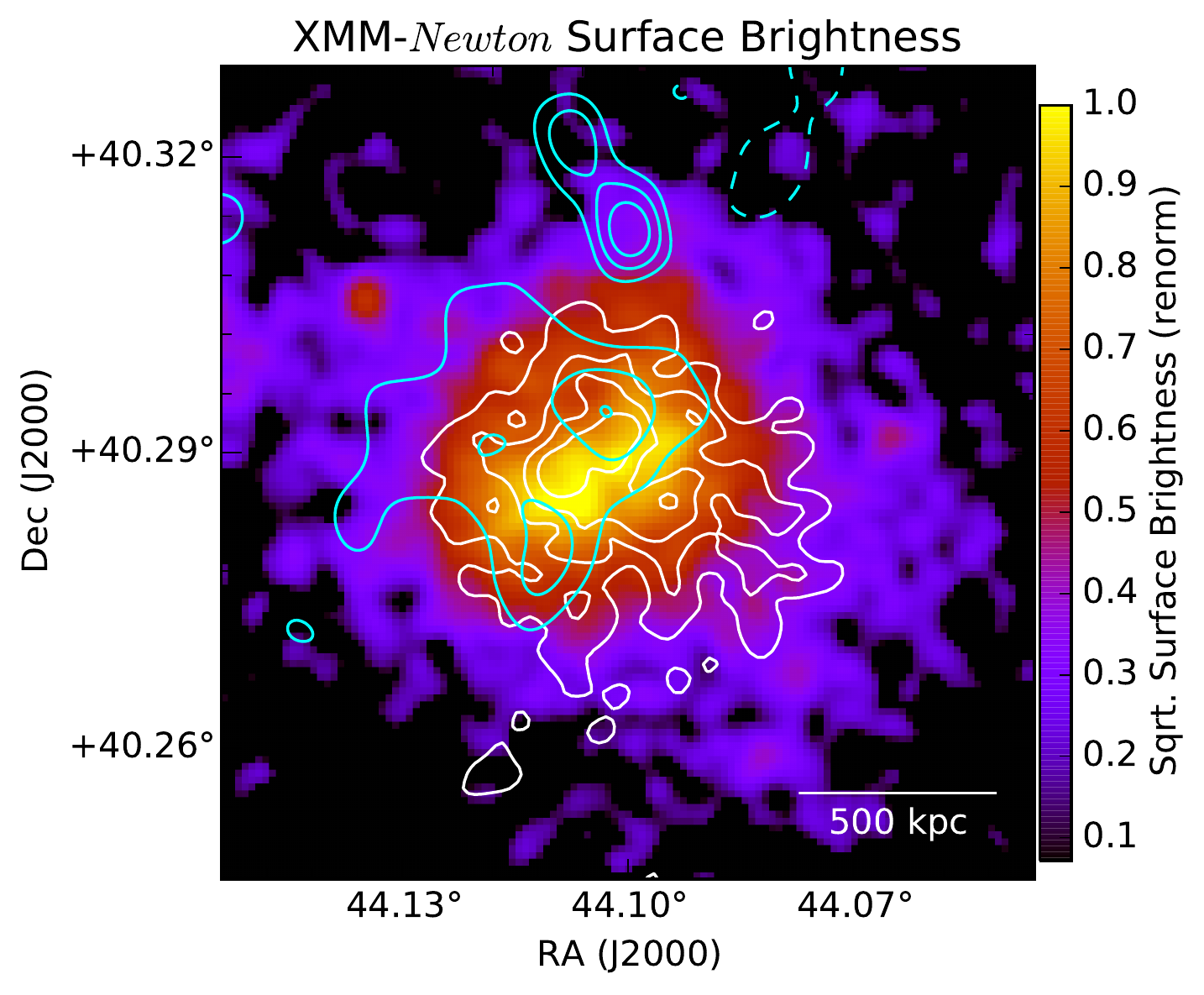}
 \includegraphics[height=1.88in]{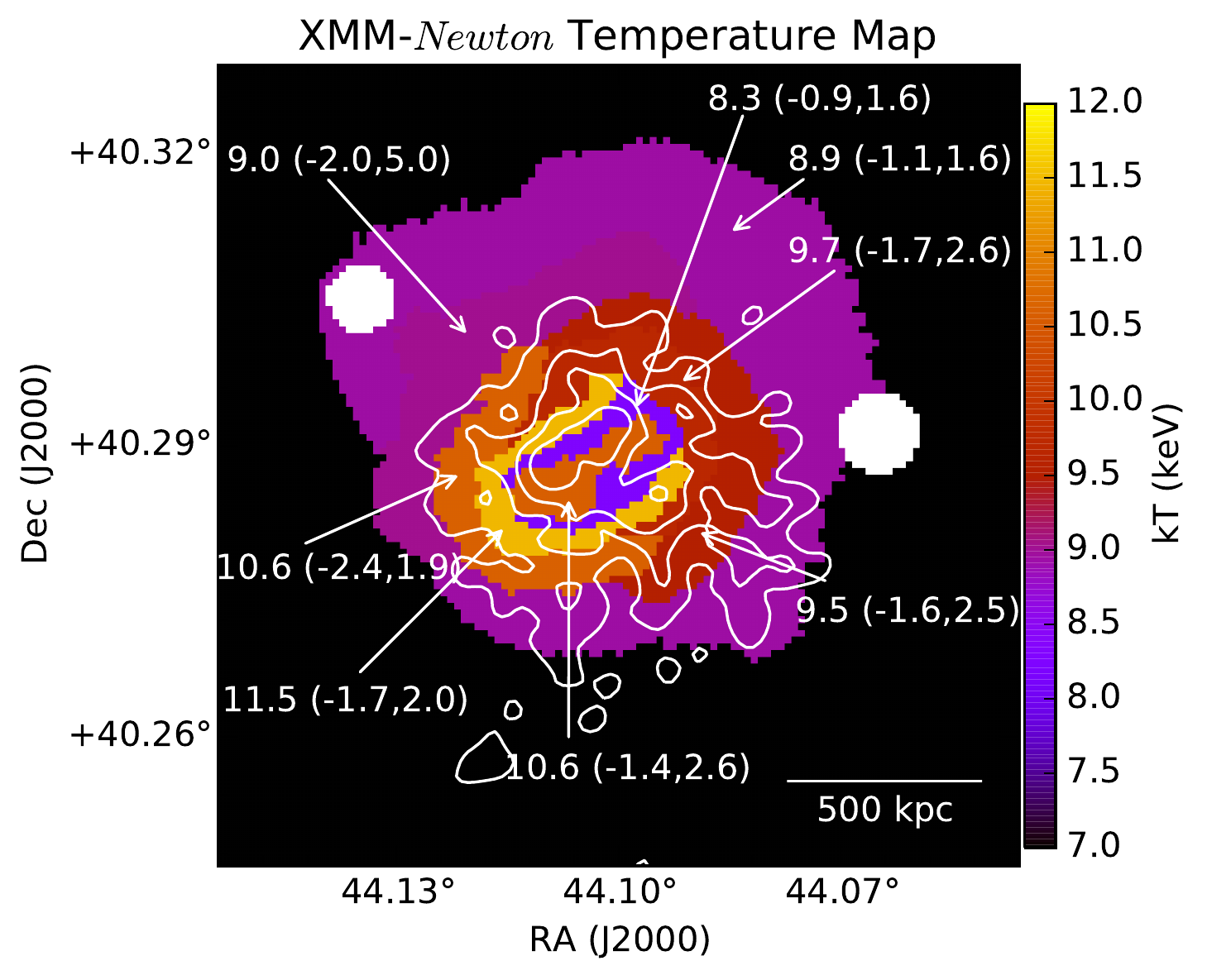}
 \includegraphics[height=1.88in]{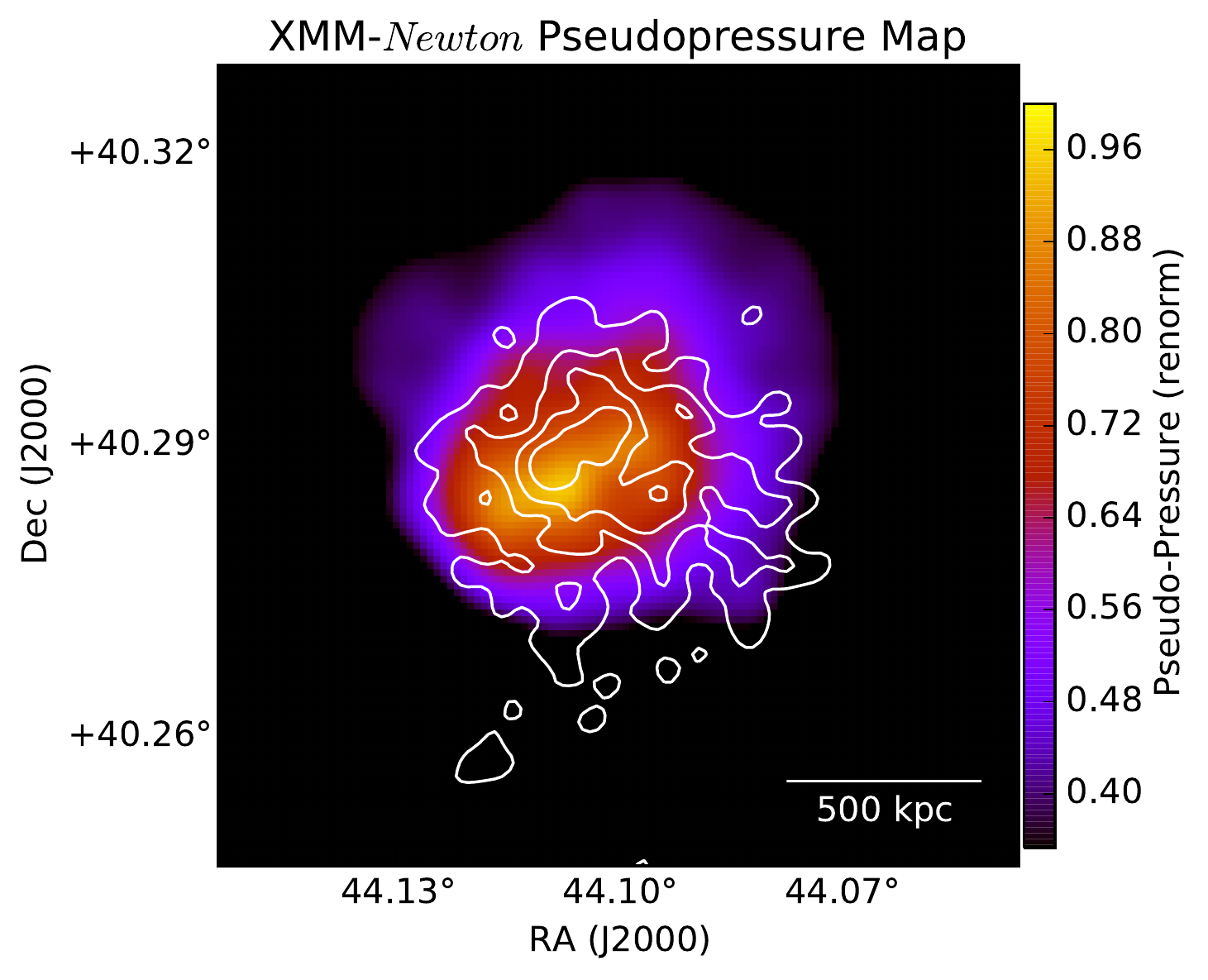}
\end{center}
\caption{ 
  {\bf Left:} Merged, background-subtracted \xmm\ \Sx\ image
  overlaid with GMRT 610~MHz contours (cyan) to match those in \cite{vanweeren2014} and with
  GISMO contours at $S/N=[-4,-6,-8,-10]$ from Figure \ref{fig:cluster} overlaid (white). 
  The central GMRT contours show the location of the radio halo, while those to the north of 
  the cluster are due two to unresolved compact radio sources.
  The X-ray image, shown on a square root scale ($\propto$ density), 
  is binned into $2.\!\arcsec5$$\times$$2.\!\arcsec5$ pixels and smoothed with 
  a 10\arcsec\ FWHM Gaussian. 
  {\bf Middle:} \xmm\ temperature map (in keV) using the contour binning 
  algorithm of \cite{sanders2006}, described in Section \ref{sec:xmm}.
  Temperature labels show 1-$\sigma$ confidence intervals in parentheses.  
  The two white regions in the map are masked due to X-ray point sources.  
  GISMO contours are depicted as in the left panel.
  {\bf Right:} Pseudo-pressure map derived by multiplying the $\sqrt{\Sx}$ image (left) 
  by the temperature map (middle panel) and smoothing to the resolution of GISMO.
  \label{fig:cluster_xray}}
\end{figure*}

\subsection{XMM-{\it Newton}}\label{sec:xmm}

The discovery of \plckg\ was confirmed through a 16.9~ksec X-ray 
observation with the \xmm\ European Photon Imaging Camera (EPIC) in August 2011, as
part of the {\it Planck} \xmm\ validation program \citep{Planck2013_XMMValidation}.
An additional 43.9~ksec \xmm\ observation was obtained in August 2012.
We calibrated both datasets using the Science Analysis System (SAS, ver.~13.5.0) 
and the most recent calibration files as of July 2014. 
The calibrated, cleaned event files discard periods of high intensity due to background 
particle flares. 
Effective exposure times and other observation details are summarized in Table \ref{tab:obs}. 

The images and spectra were cleaned of point sources. CCD4 of the MOS1 detector
was operating in an anomalous state during observation 0679181301, and therefore 
excluded from further analysis. 
The merged, exposure-corrected X-ray surface brightness (\Sx) image 
in Figure~\ref{fig:cluster_xray} (left) shows that the cluster has a disturbed 
morphology, with an elongated core angled $\approx$30$^\circ$ counterclockwise from E-W.

We extracted spectra for the temperature analysis using the SAS tools 
{\it mos-spectra} and {\it pn-spectra}.
Extracted MOS1+MOS2 spectra were co-added, and the MOS and PN spectra were
jointly fit using {\it xspec} using {\it cstat} statistics.
Each region contained $>$2000 background-subtracted (i.e.\ source) counts. 
With the column density of hydrogen held fixed at the Galactic value,
we fit the APEC plasma model to find the 
temperature of each region, marginalizing over abundance.
Further details of the data reduction and analysis such as the treatments of the 
local, cosmic, and particle backgrounds are discussed in \cite{bulbul2012}.   

We used the \xmm\ spectroscopic data to produce a temperature ($\kB\te$) map of the  
cluster using the contour binning algorithm of \cite{sanders2006}, {\it contbin}, which 
selects regions of similar \Sx\ above a user-specified $S/N$ threshold. We used $S/N>30$ 
yielding $>$2000 source counts per region.
The resulting temperature map is shown in Figure \ref{fig:cluster_xray} (middle panel), with
GISMO contours overlaid for comparison. A pseudo-pressure map ($\kB \te \times \sqrt{\Sx}$) 
is shown in Figure \ref{fig:cluster_xray} (right).

\section{Results}\label{sec:results}

In this section we compare the surface brightness of the GISMO SZ map with
the model fit to the {\it Planck} data and to the properties inferred
from the \xmm\ observations.
The thermal SZ effect traces the line-of-sight integral
of thermal electron pressure \Pe. Its surface brightness is  
proportional to the Compton $y$ parameter,
\begin{equation}
  \label{eq:compy}
  y \equiv \frac{\sigT}{\mec} \int \! \dene \kB \te \,d\ell\\
  = \frac{\sigT}{\mec} \int \! \Pe \,d\ell\\
\end{equation}
where \sigT\ is the Thomson cross-section, \kB\ is the Boltzmann constant, 
\mec is the electron rest energy, \dene\ is the electron number density, 
\kB\te\ is the electron temperature,  
$\ell$ is the line of sight path through the cluster, and $\Pe=\dene\kB\te$.

\begin{figure}
\begin{center}
	\includegraphics[width=3.4in]{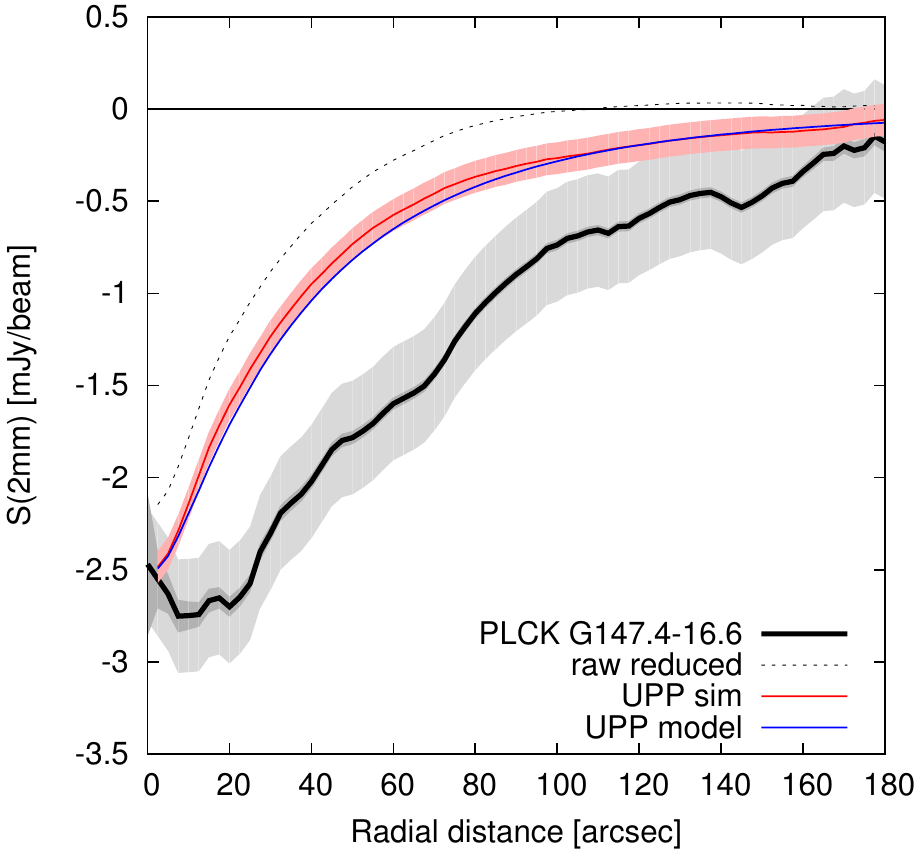}
\end{center}
\caption{SZ surface brightness profiles from the raw (dotted) and deconvolved (solid black) GISMO maps, with measurement uncertainties (dark grey) and systematic zero-level estimation uncertainty (light grey) ranges, compared with that computed from the median UPP (blue) fit to the {\it Planck} data \citep{Planck2013_XMMValidation}. We also show the profile we recover (red with 2-$\sigma$ uncertainties) when we insert the UPP into the jackknifed GISMO timestreams and analyze it the same way as the actual cluster observation. 	
	\label{fig:sz_profiles}}
\end{figure}

\cite{Planck2013_XMMValidation} report a spherically-integrated Compton 
$\Ysphf = (5.2\pm1.7)\times10^{-4}~\rm arcmin^2$, where
\begin{equation}
  \label{eq:Ysph}
  \Ysphf \equiv \frac{\sigT}{\mec \DA^2} \int_0^{R_{500}} \! \Pe(r)\, 4\pi r^2 dr.
\end{equation}
Here \DA\ is the angular diameter distance to the cluster, and $R_{500}$, which 
for this cluster $=1042~\rm kpc$ (2.\!\!\arcmin5 on the sky), is the
radius within which the average density is 500$\times$ greater than the 
critical density of the Universe at that redshift.
\cite{Planck2013_XMMValidation} assume the spherically symmetric 
`universal pressure profile' (UPP) of \citet{arnaud2010} for their model fit, shown in blue 
on Figure~\ref{fig:sz_profiles}.  For simplicity, we plotted only the median UPP fit
to the {\it Planck} data, noting the error bars on the SZ surface brightness profile 
are $\approx13\%$ at $R_{500}$, and $\approx7\%$ at the peak.

The source of the discrepancy between the measured and modelled profiles is unknown, but may 
be an indication that the UPP is a poor fit to this disturbed cluster.  
We note that the UPP as fit by the {\it Planck} is treated as a matched filter function of a 
single parameter, namely the mass within $R_{500}$ as computed using the scalings reported 
in \citet{arnaud2010}. 
Furthermore, the {\it Planck} measurement of this cluster is entirely
unresolved within $R_{500}$, so the profile shown is an interpolation of the UPP that results
in the integrated signal measurement by {\it Planck} on scales $>> R_{500}$.

The GISMO $S/N$ map shows broad qualitative agreement with the X-ray imaging
(Figure \ref{fig:cluster_xray}, left).  
In the core however, we find the SZ and X-ray peaks are offset. 
An attempt to realign the peaks would 
bring the agreement on arcminute scales into tension.

The observed peak offsets between the GISMO and \xmm\ maps is not expected 
to be due to pointing errors.  
\xmm\ imaging is precise to the subarcsecond level, and the positions of
bright X-ray point sources agree with the locations of bright WISE counterparts.  
The GISMO pointing model is checked against 466 measurements of bright point 
sources, observed hourly through several days. All are within an RMS deviation 
$<4\arcsec\!\!.5$. The GISMO observations combine 32 observation blocks, each 
bracketed by independent pointings, thus the statistical 1-$\sigma$ pointing 
error of the composite map is estimated at 0\arcsec\!\!.8.
We also note that the positions of dozens of bright compact sources observed by GISMO observations
during the same observation period are reproduced within the expected accuracy. 
We therefore consider the discrepancy to be of astrophysical nature.

We do not expect to detect any contamination by the Cosmic Infrared Background 
(CIB). The unresolved part of the CIB is removed by flux zeroing outside of 
the cluster, so our fluxes are 
effectively referenced against the mean CIB level. Resolved CIB sources may be 
present in the map, but are unlikely to be detectable. A deep-field study with GISMO by
\citet{Staguhn2014} finds no sources brighter than $\sim$1\,mJy ($>$3-$\sigma$ in our 
map) in a similar area to our field, and put the 2\,mm confusion noise at 
$\lesssim$50$\mu$Jy, i.e.\ several times below the RMS 
in the observations presented here. 
The combined CMB+CIB at scales from 1--4\arcmin\ has also been measured at
150~GHz by the SPT to be $\lesssim$100\,$\mu$K \citep{George2015}, corresponding 
to a signal $\lesssim$\,$407\mu$Jy/bm in the GISMO data, comparable to the 
noise level in our map. 
We also note that recent studies by \cite{sayers2013c} and \cite{Gralla2014} both 
found the radio and submillimeter point source contribution to be minimal near 150~GHz.
Therefore, CIB/CMB contamination in our map is expected to be small.

SZ and X-ray imaging are sensitive to the
line of sight integrals of pressure (Eq.~\ref{eq:compy}) and density squared, respectively;
for bremsstrahlung emission, X-ray surface brightness has only a weak 
dependence on temperature, $\Sx \propto \int \! \dene^2 \, \te^{1/2} d\ell$. 
Differences between the SZ and X-ray maps 
($y$ and \Sx) can therefore be due to temperature substructure or to the 
differing line of sight distribution of the gas (e.g. clumping or asphericity).

The location of the SZ signal at $>$6-$\sigma$ broadly 
agrees with the location of the hottest gas found in the temperature and
 pseudo-pressure maps (Figure \ref{fig:cluster_xray}, 
middle and right, respectively),
but we find no clear evidence for shock-heated gas at the resolution of the X-ray spectroscopy. 
For the high-significance ($>$6-$\sigma$) SZ region, we extracted spectra from both \xmm\ 
observations, and X-ray counts within $R_{500}$, using the same fitting procedure and 
plasma model as for deriving the temperature. We find only a marginal enhancement of
$\Tx=11.33_{-1.61}^{+2.35}$~keV over the global temperature 
$\Tx_{,500}=8.74_{-0.56}^{+0.58}$~keV, which agrees with that found 
by \citet{Planck2013_XMMValidation}.

This leaves the possibility that the SZ/X-ray offset is due to an irregular
gas distribution along the line of sight or the breakdown of the assumption
that pseudo-pressure and SZ features should directly match.  
The so-called `slab approximation', which treats the line of sight temperature as isothermal in
each spectroscopic bin and assumes the path length through the cluster is a constant
\citep[e.g][]{mroczkowski2012,planck2013x}, may not hold for complicated merger geometries.

\section{Conclusions}\label{sec:conclusions}

We present the first high significance maps of the SZ effect with GISMO, revealing
substructure in the {\it Planck}-selected cluster \plckg.  
The core morphology mimics the appearance of the X-ray 
observation \citep[reported here and in][]{Planck2013_XMMValidation}, 
but is notably offset from their X-ray counterparts.   
The presence of the giant radio halo reported in \citealt{vanweeren2014} 
further supports the hypothesis that this system is likely
a merger.

This GISMO observation demonstrates that a comparable level of detail in a 
moderately high-$z$ cluster can now be obtained from large, ground-based 
telescopes in a similar amount of time as that currently required for X-ray 
observations.  This adds to a small but growing number of instruments that have imaged
the SZ effect at resolutions better than 20\arcsec, which include
such instruments as Nobeyama \citep[e.g.][]{komatsu2001,kitayama2004}, 
MUSTANG \citep[e.g.][]{mason2010}, CARMA \citep[e.g.][]{plagge2010}, and NIKA 
\citep[e.g.][]{Adam2014,Adam2015}.
New and future observations with GISMO and GISMO-2 \citep{Staguhn2012} 
will probe high-$z$ cluster mergers further, or confirm cluster candidates 
from SZ, X-ray, and optical surveys.

\acknowledgments

We thank all of the staff at the IRAM 30-meter for their support, 
and Israel Hermelo in particular. 
We also thank Rafael Eufr{\'a}sio and Alexander Karim for their input
on optimizing observational strategies with GISMO+IRAM, and we
thank the anonymous referee for the diligence that led to vast
improvements in our analysis of the GISMO SZ data.
And finally, we thank Monique Arnaud for her insightful comments on the 
\xmm\ observations.

This research was performed while TM held a National Research Council 
Research Associateship Award at the Naval Research Laboratory (NRL).
Basic research in radio astronomy at NRL by TM and TEC is supported by 6.1 Base funding.
EB is supported in part by NASA ADP grant NNX13AE83G.
RJvW is supported by NASA through the Einstein Postdoctoral
grant number PF2-130104 awarded by the {\it Chandra} X-ray Center, which is
operated by the Smithsonian Astrophysical Observatory for NASA under
contract NAS8-03060.
HTI is supported by the National Radio Astronomy Observatory, a
facility of the National Science Foundation (NSF) operated under cooperative
agreement by Associated Universities, Inc.
The GISMO instrument and team are supported through NSF ATI grants 
1020981 and 1106284. 
IRAM is supported by INSU/CNRS (France), MPG (Germany) and IGN (Spain).

\defcitealias{Planck2011X}{Planck Collaboration X 2011}
\defcitealias{Planck2011XI}{Planck Collaboration XI 2011}
\defcitealias{Planck2011XII}{Planck Collaboration XII 2011}
\defcitealias{Planck2013XXIX}{Planck Collaboration XXIX 2014}
\defcitealias{Planck2013XXIX_update}{Planck Collaboration XXIX Update 2015}
\defcitealias{Planck2015XXVII}{Planck Collaboration XXVII 2015}
\defcitealias{Planck2013v}{Planck Collaboration Intermediate Results V 2013}
\defcitealias{Planck2013x}{Planck Collaboration Intermediate Results X 2013}


\end{document}